\documentclass[runningheads]{llncs}
\usepackage{graphicx}
\usepackage{algorithm}
\usepackage{algpseudocode}
\begin{document}
\title{ArxNet Model and Data: Building Social Networks from Image Archives}
%
%
\author{Haley Seaward \and
Jasmine Talley \and
David Beskow}
\authorrunning{H. Seaward et al.}
%
\institute{United States Military Academy, West Point NY 10996, USA 
\email{david.beskow@westpoint.edu}\\
}
\maketitle              
\begin{abstract}
A corresponding explosion in digital images has accompanied the rapid adoption of mobile technology around the world.  People and their activities are routinely captured in digital image and video files.  By their very nature, these images and videos often portray social and professional connections.  Individuals in the same picture are often connected in some meaningful way.  Our research seeks to identify and model social connections found in images using modern face detection technology and social network analysis.  The proposed methods are then demonstrated on the public image repository associated with the 2022 Emmy's Award Presentation.

\end{abstract}
\section{Introduction and Background}\label{sec1}

The increased use of photography in the 20th century led to the phrase ``a picture is worth a thousand words,'' first recorded used circa 1911.  Images contain a wealth of information. Images of people portray emotion, personality, economic status, and other information about the people and events that are etched into pixels.  Pictures also record human relationships.  Images that contain more than one person often indicate some level of connection or relationship among the individuals captured in a still photo or motion images.  These could be family, business, friendship, acquaintance, or other types of social ties.  These connections can be captured if we have a method to identify the individuals in the image.  This capability came in the form of Face Detection.  While face detection research had an early start in US intelligence organizations in the 1960s \cite{de2007history}, it became usable in the late 1990s and scientists now had the method to identify unique individuals in images.  With face detection technology in hand, some past research has focused on building social networks from a family picture archive \cite{zhang2003automated} and from movies \cite{weng2009rolenet}.  Our research will focus on building these networks on event data using modern open-source facial detection software.  Additionally, our algorithms become computationally scalable by incorporating efficient search algorithms.   This algorithm is demonstrated on a specific public event, the 2022 Emmy's Awards held on 12 September 2022.  The algorithm as well as the face detection embeddings and resulting social network will be made public once this paper is published.  

\section{Previous Work}\label{sec2}

Face detection is one of the most studied aspects of computer vision.  Face detection is used for a variety of different use cases and is the first step in most human-computer and human-robot interaction \cite{zafeiriou2015survey}.  Computer vision-based faced detection began in the 1960s \cite{bledsoe1965man} but wasn't practical on normal everyday photos (photos ``in the wild'') until the Viola-Jones algorithm using boosting methods was developed in the late 1990s and early 2000s \cite{viola2001rapid}.

Face recognition involves several steps, each of which have a number of different algorithms to address.  Facial recognition generally involves the following steps:

\begin{enumerate}
    \item Face detection (finding and segmenting the face(s) in the image)
    \item Face warping (given that faces can be at an infinite number of angles, this step identifies facial landmarks and `warps' the head so that the eyes, nose, and mouth are centered)
    \item Encode the faces (create a mathematical representation of the face)
    \item Compare faces (recognize person by comparing encodings)
\end{enumerate}

With the growth of mobile technology today, image archives can easily contain millions of images.  Using a brute force method for searching face embeddings becomes computationally intractable, as we prove below.  Several methods have been developed to improve search.  These include the k-dimensional tree (KD-tree) \cite{bentley1975multidimensional}, which is successful at low-dimension embeddings.  Ball-tree \cite{omohundro1989five} and Locality Sensitive Hashing (LSH) \cite{indyk1998approximate} both work effectively at higher dimensions.  Other methods include Approximate Nearest Neighbors (ANN) \cite{indyk1998approximate} which sacrifices some accuracy for speed \cite{arya1998optimal}.  We used the CPU version of FAISS \cite{johnson2019billion} and explored the use of both the default Flat L2 Search (also a brute-force method) and the Inverted File with Product Quantization (IVFPQ) indexing.  The inverted file indexing creates inverted lists where each list contains vectors that are close to a centroid and are efficient to search.  This also uses product quantization as a compression technique that reduces the size of the index in memory, increasing the allowable index size.

Relatively few authors have looked at building relationship networks from images. In 2003 Zhang et al used a Bayesian framework to annotate faces and create a network from a family photo album by estimating face similarity, working with missing features, and generating names for unlabeled faces by comparing their similarities to a list of labeled faces \cite{zhang2003automated}. Later, Weng, Chu, and Wu created `RoleNet', an algorithm that would automatically identify movie scenes and create connections between the actors/characters that were represented in that scene \cite{weng2009rolenet}.  Zhang, Luo, and Loy expanded on this to describe interpersonal relationships from the facial expressions of face images in the wild \cite{zhang2018facial}.  This research was expanded on with networks clustering in 2020 \cite{kulshreshtha2020dynamic}.

We used the  Python face-recognition package created by Adam Geitgey to conduct facial recognition \cite{geitgey2016face_recognition}.  Our pipeline used Histogram of Oriented Gradients (HOG) \cite{dalal2005histograms} for face detection, face landmark orientation for face warping \cite{kazemi2014one}, and Open Face \cite{amos2016openface} for facial encoding.  We then used euclidean distance for face recognition given the face emebedding.  

Our research aims to build a relationship network of faces from a collection of photos from a specific event. From this relationship network, network science techniques such as centrality and clustering are used to understand the strength of the resulting relationship network.

\section{Data}\label{sec3}

\begin{figure}[h]
    \centering
    \includegraphics[width=\textwidth]{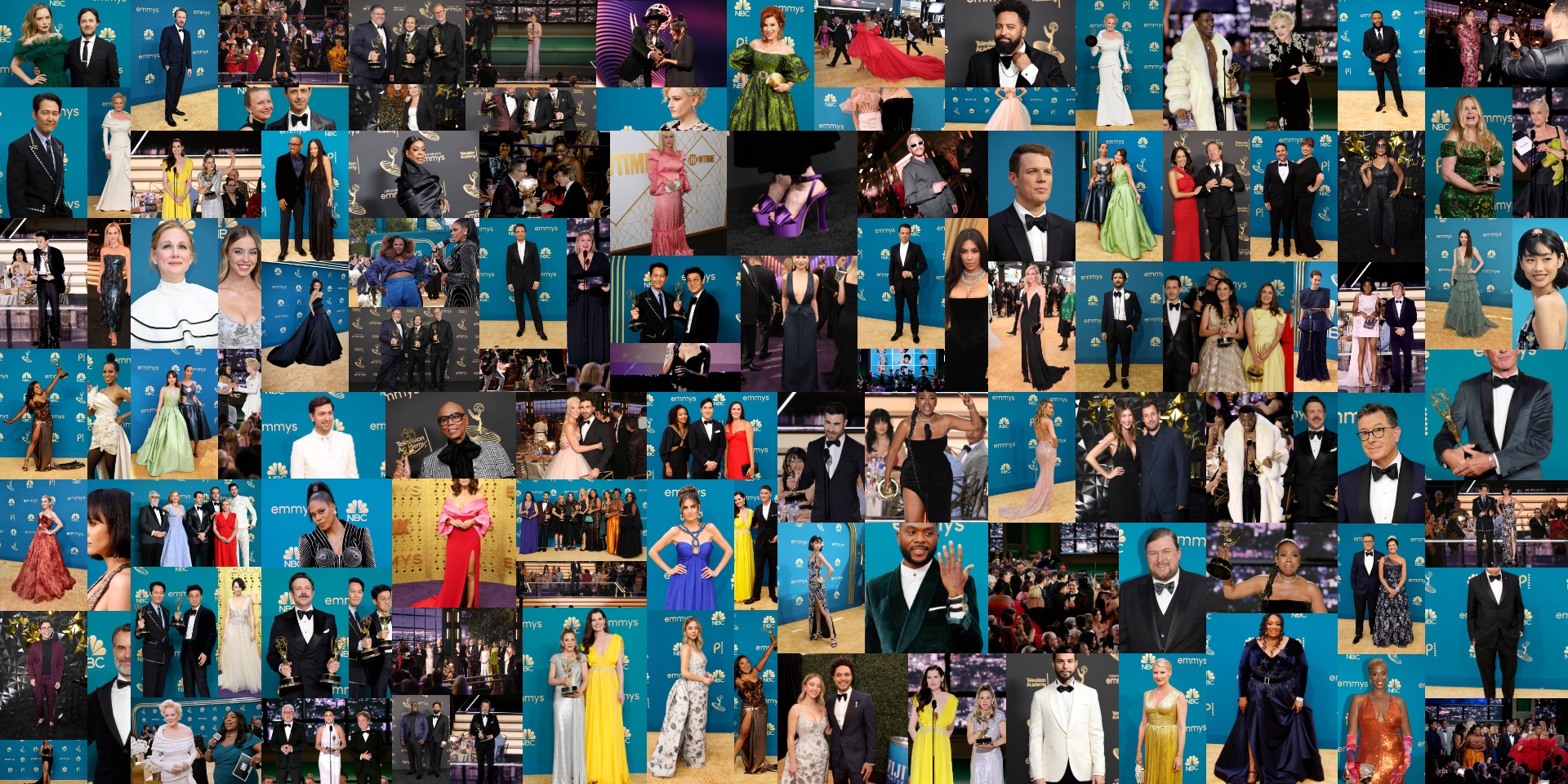}
    \caption{Example Getty images from the 2022 Emmy Award Show}
    \label{fig:collage}
\end{figure}

Our research will use the pictures from the Emmy Award Show hosted in downtown Los Angeles and held on Sept 12, 2022. These images were publicly available on Getty Images (\url{https://www.gettyimages.com/}). There are a total of 2,828 images from this event with 1,072 unique faces. A sample of these images is displayed in Figure \ref{fig:collage}.

The Emmy's Awards provides a well-known event with known celebrity faces and known celebrity links or connections.  These connections are often generated by co-starring in past or present television shows.  This resulting social network will be released for open source once this paper is accepted/published.

\section{Methods}\label{sec4}

\begin{figure}[h]
    \centering
    \includegraphics[width=\textwidth]{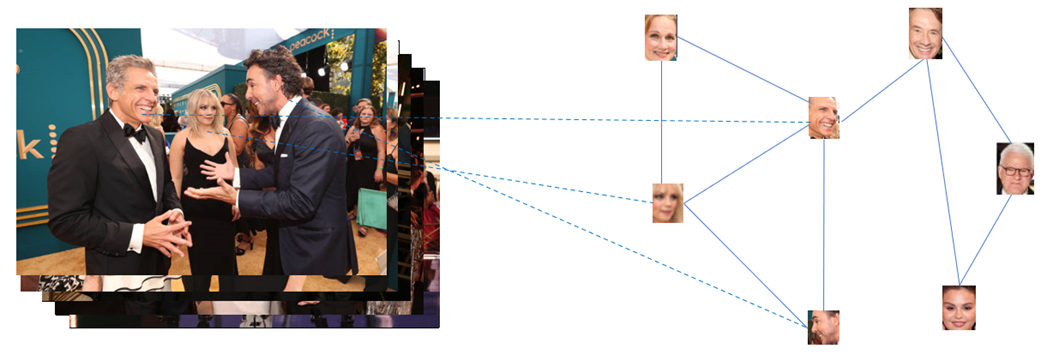}
    \caption{Illustrating the ArxNet Approach}
    \label{fig:img1}
\end{figure}

Given an archive of photos associated with a specific time, place, and event, the ArxNet approach identifies unique faces in the images and then constructs a network of co-occurring faces.  This approach is illustrated in Figure \ref{fig:img1}.  To conduct this approach, we 1) conduct face recognition to extract all face embeddings for each image in the archive, then 2) build an index for the unique faces that we find, 3) build the edge list for unique co-occurring faces, and finally 4) construct the graph from the edge list.  


The example relationship network in Figure \ref{fig:img1} begins with a photo of Ben Stiller, his daughter, Ella Stiller, and Shawn Levy. The next photo includes both Ben Stiller and his daughter again and introduces Laura Linney. The third photo once again includes Ben Stiller and Ella Stiller, and introduces Martin Short. The relationship diagram then follows Martin Short to the next photo which includes Selena Gomez, Steve Martin, and of course Martin Short. The last photo in this relationship diagram includes the same three faces, but then adds a large crowd in the background. The identifiable faces in the background of the last photo are then also included in the relationship network. This example relationship network is simply a digestible snapshot of the larger Gephi Relationship diagram which covers all the Getty Images from the 2022 Emmy Award Show. 

Face recognition was conducted with Adam Geitgey's pipeline \cite{geitgey2016face_recognition}.  We modified his Face Recognition Docker Image to add compatibility with a Jupyter environment.  This pipeline uses Histogram of Oriented Gradients (HOG) \cite{dalal2005histograms} for face detection, face landmark orientation for face warping \cite{kazemi2014one}, and OpenFace \cite{amos2016openface} for facial encoding.  At the end of this pipeline, each face is represented with an embedding of length 128 that we store in a data structure where each image name is associated with a list of the face embeddings found in it.  

Face recognition was conducted using the euclidean distance between two face embeddings.  If the distance between the embeddings was less than 0.5, then the embeddings were deemed to be the same face.  While we conducted some exploration of the right distance threshold, we found the default setting of 0.5 to be adequate for our use case.  

Using the resulting data structure, we developed an algorithm that would build the face co-occurrence network.  This basic algorithmic approach is provided in Algorithm \ref{alg1}.  This algorithm takes each face for each picture, and first searches an index to see if we've already seen the face before.  If found, we get the face indices and associate it with the picture.  If the face is not found, we add the face to our index.  Once we've identified all of the faces in the image, then we can associate them with a co-occurrence link.  To do this we identify all combinations of two faces from the set of faces in the picture.  For each combination, we add a link or edge to an edge list.  



\begin{algorithm}[htb]
    \caption{Basic ArxNet Algorithmic Approach}
    \label{alg1}
    \begin{algorithmic}
        \State initialize index
        \For{each picture}
            \For{each face}
                \State search index
                \If{found}
                    \State get key
                \Else 
                    \State add face to index
                \EndIf
            \EndFor
            \If{faces in picture $> 1$}
                \State calculate all combinations of 2
                \For{each combination}
                    \State append to edgelist
                \EndFor
            \EndIf
        \EndFor
    \end{algorithmic}
\end{algorithm}

\begin{figure}[htb]
    \centering
    \includegraphics[width=\textwidth]{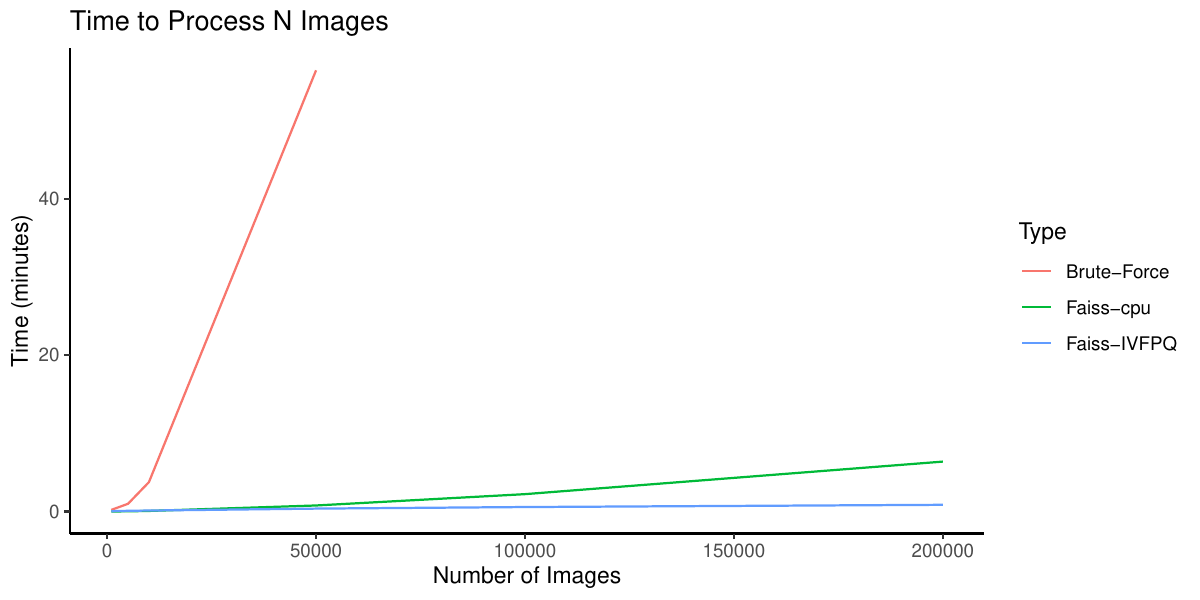}
    \caption{Demonstrating computational time complexity of different search index}
    \label{fig:faiss}
\end{figure}

For our initial algorithm for the Emmy Award data, we used a brute force search algorithm in base Python with time complexity of $O(n^2)$. For the Getty Images dataset this ran in approximately 40 seconds on a standard laptop.  This method did not scale, as illustrated in Figure \ref{fig:faiss}.  In order to improve the search speed, we leveraged the FAISS package created by the Facebook Team \cite{johnson2019billion}.  We used the CPU version of FAISS and explored the use of both the default Flat L2 Search (also a brute-force method) and the Inverted File with Product Quantization (IVFPQ) indexing.  The inverted file indexing creates inverted lists where each list contains vectors that are close to a centroid and are efficient to search.  The increased speed of these is clear in Figure \ref{fig:faiss}.  Using this method we were able to run the ArxNet Algorithm on 200K images in approximately 50 seconds.  

\section{Results}\label{sec5}

The algorithm generates an edgelist and network representing co-occurring faces. The resulting network reveals that there are 1072 unique faces, of which 941 become nodes since they have relationships with at least one other faces. The 941 nodes have a total of 3726 edges which connect the nodes together of the Sept 12, 2022 Emmy Award show hosted in Downtown, Los Angeles.  This network contains 88 unique communities, of which 55 only contain two nodes (or faces).  The largest community contains 10\%  of the nodes (98/941 nodes).

\begin{figure}[h]
    \centering
    \includegraphics[width = 0.7\textwidth]{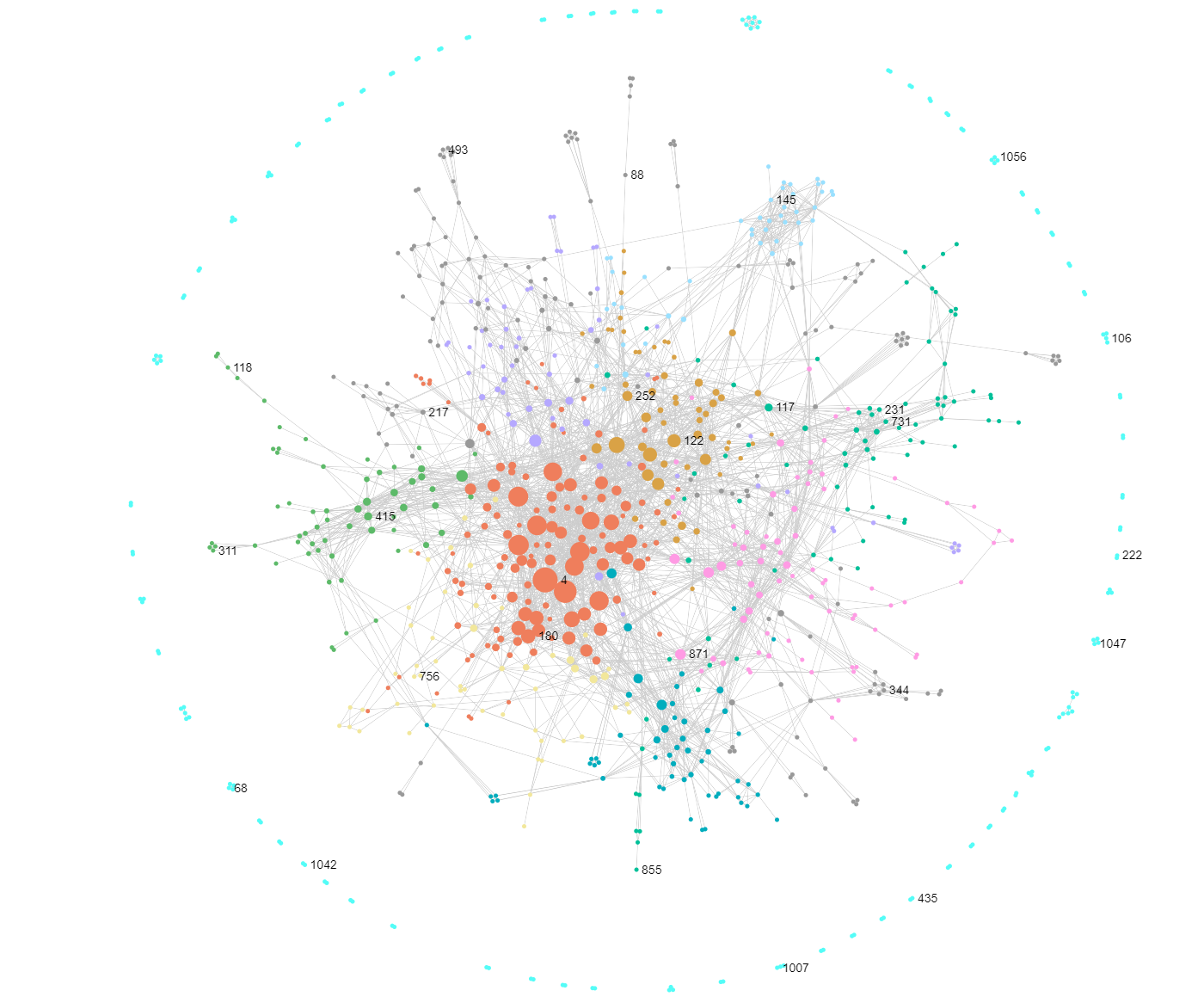}
    \caption{Emmy's ArxNet Relationship Diagram (colored by Louvain Cluster)}
    \label{fig:img4}
\end{figure}

Another way to look at the diagram is by partitioning the diagram by edges and then specifically through the attribute ‘images’. When the diagram is partitioned this way, the image with the largest number of edges is Figure 2. This image also comes from community 41 and is responsible for 2.07\% of the total edges in the relationship network, or 77/3726 edges. 

\begin{figure}[h]
    \centering
    \includegraphics[width = 0.6\textwidth]{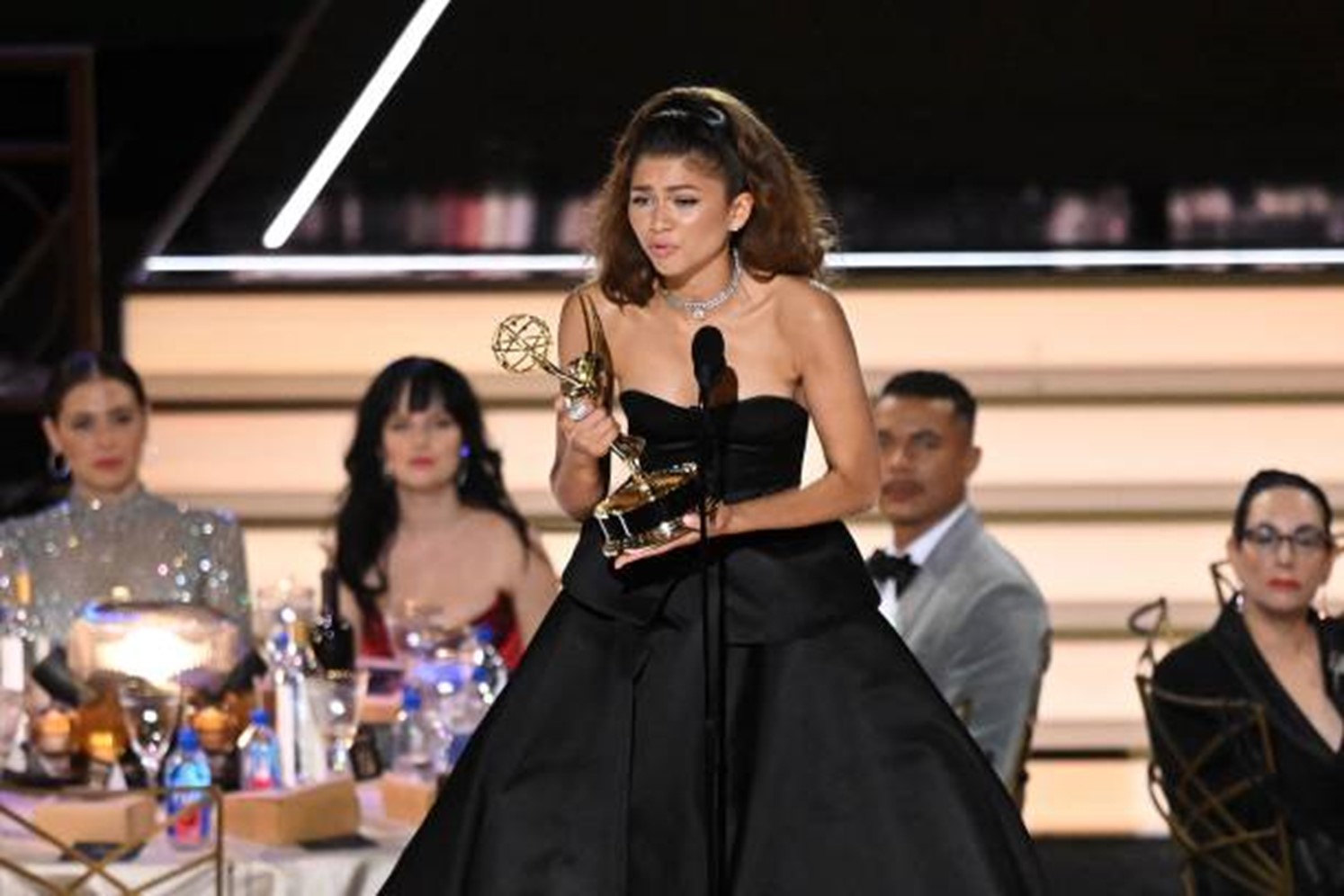}
    \caption{Image with most edges}
    \label{fig:img2}
\end{figure}

Figure \ref{fig:img2} depicts actress, Zendaya accepting the Emmy for Outstanding Lead Actress in a Drama Series for the television series “Euphoria”. The most likely reason this photo is responsible for so many edges is because of its location. There are numerous photos that are taken in this spot because this is the spot in which Emmy award winners address the audience after winning an award. Therefore, the people in the background of this photo are also in the background of many other photos taken in this same location. Although these people are in the background of numerous photos, it is still noteworthy that Zendaya is the fifth person in this photo. Out of all the photos with these people in the background, the photo with Zendaya is responsible for the most edges. This reveals that Zendaya along with the four people in the background share the most relationships with other faces in other photos.

\section{Conclusion and Future Work}\label{sec6}
In this paper, we have addressed the previous work in the fields of face recognition, search, and building community networks from image archives. We have presented a technique that identifies faces via face embeddings, creates relationships, and then groups relationships into communities. We validated a threshold for the Euclidean distance threshold for determining the same face. After validating the threshold, we built the relationship network. The network connected 941 attendees of the 2022 Emmy Award Show. The output also revealed the strongest communities and the single image that is responsible for the most edges in the relationship network. 

The limiting factor to this approach is that the algorithm creates relationships between people in the foreground and the background of the photos. This limitation is prevalent in the photo with the strongest edges. This photo has four people in the background and one person in the foreground. This skews the results because the algorithm gives equal weight to relationships that consist of both faces in the foreground and relationships that consist of both foreground and background faces. It is possible that the people in the background and the foreground do have relationships, but it is not fair to assume that these relationships are as strong as relationships that are both foreground and backgrounds.

Social connections provide the foundation of social and organizational interactions and evolution.  Mapping these connections through facial co-location in image archives can enable the study, understanding, and mapping of these critical connections.

\bibliographystyle{splncs04}
\bibliography{our_paper}

\begin{thebibliography}{10}
\providecommand{\url}[1]{\texttt{#1}}
\providecommand{\urlprefix}{URL }
\providecommand{\doi}[1]{https://doi.org/#1}

\bibitem{amos2016openface}
Amos, B., Ludwiczuk, B., Satyanarayanan, M.: Openface: A general-purpose face
  recognition library with mobile applications. Tech. Rep. CMU-CS-16-118, CMU
  School of Computer Science (2016)

\bibitem{arya1998optimal}
Arya, S., Mount, D.M., Netanyahu, N.S., Silverman, R., Wu, A.Y.: Optimal
  approximate nearest neighbors in high-dimensional spaces. Journal of the ACM
  (JACM)  \textbf{45}(6),  891--923 (1998)

\bibitem{bentley1975multidimensional}
Bentley, J.L.: Multidimensional binary search trees used for associative
  searching. In: Communications of the ACM. vol.~18, pp. 509--517. ACM (1975)

\bibitem{bledsoe1965man}
Bledsoe, W.W., Chan, H.: A man-machine facial recognition system—some
  preliminary results. Panoramic Research, Inc, Palo Alto, California.,
  Technical Report PRI A  \textbf{19}, ~1965 (1965)

\bibitem{dalal2005histograms}
Dalal, N., Triggs, B.: Histograms of oriented gradients for human detection.
  In: Proceedings of the IEEE Computer Society Conference on Computer Vision
  and Pattern Recognition (CVPR). pp. 886--893. IEEE (2005)

\bibitem{geitgey2016face_recognition}
Geitgey, A.: face\_recognition: A simple face recognition library for python.
  \url{https://github.com/ageitgey/face\_recognition} (2016)

\bibitem{indyk1998approximate}
Indyk, P., Motwani, R.: Approximate nearest neighbors: towards removing the
  curse of dimensionality. Proceedings of the thirtieth annual ACM symposium on
  Theory of computing pp. 604--613 (1998)

\bibitem{johnson2019billion}
Johnson, J., Douze, M., J{\'e}gou, H.: Billion-scale similarity search with
  {GPUs}. IEEE Transactions on Big Data  \textbf{7}(3),  535--547 (2019)

\bibitem{kazemi2014one}
Kazemi, V., Sullivan, J.: One millisecond face alignment with an ensemble of
  regression trees. In: Proceedings of the IEEE conference on computer vision
  and pattern recognition. pp. 1867--1874 (2014)

\bibitem{kulshreshtha2020dynamic}
Kulshreshtha, P., Guha, T.: Dynamic character graph via online face clustering
  for movie analysis. Multimedia Tools and Applications  \textbf{79},
  33103--33118 (2020)

\bibitem{de2007history}
de~Leeuw, K.M.M., Bergstra, J.: The history of information security: a
  comprehensive handbook. Elsevier (2007)

\bibitem{omohundro1989five}
Omohundro, S.M.: Five balltree construction algorithms. In: International
  Computer Science Institute Technical Report. vol. TR-89-063 (1989)

\bibitem{viola2001rapid}
Viola, P., Jones, M.: Rapid object detection using a boosted cascade of simple
  features. In: Proceedings of the 2001 IEEE computer society conference on
  computer vision and pattern recognition. CVPR 2001. vol.~1, pp.~I--I. Ieee
  (2001)

\bibitem{weng2009rolenet}
Weng, C.Y., Chu, W.T., Wu, J.L.: Rolenet: Movie analysis from the perspective
  of social networks. IEEE Transactions on Multimedia  \textbf{11}(2),
  256--271 (2009)

\bibitem{zafeiriou2015survey}
Zafeiriou, S., Zhang, C., Zhang, Z.: A survey on face detection in the wild:
  past, present and future. Computer Vision and Image Understanding
  \textbf{138},  1--24 (2015)

\bibitem{zhang2003automated}
Zhang, L., Chen, L., Li, M., Zhang, H.: Automated annotation of human faces in
  family albums. In: Proceedings of the eleventh ACM international conference
  on Multimedia. pp. 355--358 (2003)

\bibitem{zhang2018facial}
Zhang, Z., Luo, P., Loy, C.C., Tang, X.: From facial expression recognition to
  interpersonal relation prediction. International Journal of Computer Vision
  \textbf{126},  550--569 (2018)

\end{thebibliography}
\end{document}